\documentclass[prb,twocolumn,showpacs,nofootinbib]{revtex4}
\usepackage[dvips]{graphicx}
\usepackage{color,array,dcolumn}
\usepackage{amsmath}
\usepackage{amssymb}

\begin{document}

\title{Variational Monte Carlo for spin-orbit interacting systems}
\author{A. Ambrosetti}
\email{ambroset@pd.infn.it}
\affiliation{Dipartimento di Fisica, University of Padova, via Marzolo 8, I--35131, 
Padova, Italy}
\author{P.L. Silvestrelli}
\affiliation{Dipartimento di Fisica, University of Padova, via Marzolo 8, I--35131, 
Padova, Italy}
\author{F. Toigo}
\affiliation{Dipartimento di Fisica, University of Padova, via Marzolo 8, I--35131, 
Padova, Italy}
\author{L. Mitas}
\affiliation{Department of Physics, North Carolina State University, Raleigh, NC 27695-8202, USA}
\author{F. Pederiva}
\affiliation{Dipartimento di Fisica, University of Trento, via Sommarive 14, I--38123, 
Povo, Trento, Italy}
\affiliation{INFN, Gruppo Collegato di Trento, Trento, Italy}

\begin{abstract}

\date{\today}
Recently, a diffusion Monte Carlo algorithm was applied to the study of spin dependent
interactions in condensed matter \cite{io}. 
Following some of the ideas presented therein, and applied to a Hamiltonian containing 
a Rashba-like interaction, a general variational Monte Carlo
approach is here introduced that treats in an efficient and very accurate way the spin degrees of freedom
in atoms when spin orbit effects are included in the Hamiltonian describing the electronic structure.
We illustrate the algorithm on the evaluation
of the spin-orbit splittings of isolated C, Tl, Pb, Bi, and Po atoms. In the case of the carbon atom,
we investigate the differences between the inclusion of spin-orbit in its realistic and effective 
spherically symmetrized forms. 
The method exhibits a very good accuracy in describing the small energy 
splittings, opening the way for a systematic quantum Monte Carlo studies of spin-orbit effects in
atomic systems.
\end{abstract}

\maketitle

\section{Introduction}
In the last few decades, Quantum Monte Carlo (QMC) techniques were successfully applied to a 
large number of systems in different fields of physics, quantum chemistry 
and materials science \cite{cep1,mitasrev,cep2}.
The high accuracy obtained and the possibility of treating a relatively large number of
particles within an affordable computational cost have certainly been the key ingredients for
such a success.
A fundamental property of QMC methods resides also in their scalability, which, due to
the introduction and recent proliferation of massively parallel systems,  makes a good case
for a future applicability to large quantum many-body systems, up to now mainly investigated
by means of Density Functional methods.
 Several years ago, an extension of the diffusion Monte Carlo (DMC) algorithm,
employing an efficient
recasting of the Green's function, was applied to many nucleon systems with
interactions depending on the spin and isospin degrees of freedom\cite{fanschm}.
 This seminal paper has opened the way to studies of broad
classes of systems in which spin related effects play a fundamental role. Most successful applications
so far attain the field of nuclear physics\cite{gandolfi}. 
However, in the last few years the method was extended to many electron systems, and in particular to 
the study of the two-dimensional homogeneous electron gas\cite{io,ioepl} and parabolic quantum dots\cite{iodot}.
These models are often used to describe quasi two-dimensional nanostructures built on semiconductor heterojunctions,
where the confining potential shows an asymmetry giving rise to a transverse electric field interacting
with the moving electrons. This coupling can be described by an effective Rashba spin-orbit Hamiltonian.
The next natural extension of the method in condensed matter application is the study of
spin-orbit (SO) effects in atoms, molecules and solids. Preliminary estimations show
that QMC methods should be able to provide the necessary accuracy and affordability 
for bringing to a completely new level the theoretical investigation in this area. 
The extension is highly non trivial, mostly due to the technical issues that might in 
principle limit the accuracy of a QMC calculation in presence of a non-central interaction.

One of the main issues concerning the DMC method is the need for trial wave functions,
i.e. accurate approximations to the exact solution of the Schroedinger's equation.
The availability of a good trial wave function represents a crucial challenge both at the
computational and at a more fundamental level. The trial
wave function not only affects the computational efficiency of the DMC calculation, but 
 for Fermionic systems, determines the accuracy of the
 final result as a consequence of the necessity of applying the so-called ``fixed-node``
 approximation''\cite{mitas2,kalos,umrigar1}.
An additional problem which comes to the forefront in the case of spin-orbit interaction
is the necessity
to deal with operators which do not necessarily commute with the rest of the Hamiltonian.

Both these problems are usually tackled via a preliminary variational Monte Carlo (VMC) approach.
Regarding the first challenge, VMC is an important, effective  
way of dealing with optimization of trial wave functions including almost arbitrarily complex 
many-body correlations\cite{umrigar2,sorella}. 
Concerning the second issue, the DMC expectations of operators (denoted as $O$)
 not commuting with the Hamiltonian is known to be biased by errors depending on the trial wave
 function accuracy as well, so that efficient optimization is again an important way to alleviate
this problem. In DMC, such errors stem from the way DMC expectation values are computed: 
\begin{equation}
<O>_{DMC}=\frac{<\phi_T|O|\psi>}{<\phi_T|\psi>}.
\end{equation}
Here $\phi_T$ represents the trial wave function while $\psi$ is the
DMC ground state of the system.
If $[O,H]=0$ this in principle introduces no bias since $\psi$ is an eigenstate of $H$.
If this is not the case,
 the estimate is affected by a bias depending on $\phi_T$, and it is usually corrected to
the leading order by using the formula
\begin{equation}
<\psi|O|\psi> \sim 2 <\phi_T|O|\psi> - <\phi_T|O|\phi_T> ,\\
\end{equation}
where integrals are normalized and the last term corresponds to the VMC estimate of
 the operator $O$ over $\phi_T$.
\newline
From these arguments it is possible to grasp how an efficient VMC algorithm capable of explicitly 
dealing with spin degrees of freedom is of significant
importance, not only as a robust and reliable numerical method itself, but also as for
the development of the more sophisticated QMC algorithms. 

In this paper we  present the theory, the Monte Carlo algorithm, and the results 
assessing the possibility of developing wave functions containing the necessary
correlation at the {\it antisymmetric} level (i.e. in the part of the wave function which
is usually described by a Slater determinant) by means of a variational procedure.
The many body correlations that are usually introduced in order to describe the
short range effects of the Coulomb interaction are neglected on purpose for the C atom in order to
make the peculiar aspects of the spin-dependent algorithm clear. 
Two body correlations are instead introduced in calculations relative to the more relevant cases of 
heavier atoms.
The paper is organized as follows. In section II the VMC algorithm and the structure of the
spinorial wave functions used in the calculations are illustrated. Section III presents applications
to the evaluation of spin-orbit splittings in C, Tl, Pb, Bi and Po.
Section IV is devoted to conclusions.

\section{Method}
Variational Monte Carlo is a very efficient algorithm for evaluating expectation values of observables over
a given trial wave function through the computation of integrals of the kind
\begin{equation}
\frac{<\psi_T|O|\psi_T>}{<\psi_T|\psi_T>}.
\end{equation}
When one restricts to Hamiltonians or operators with no explicit spin dependence, it is not necessary to perform
the summations over the spin degrees of
freedom. These become static variables, and can simply be treated as labels.
As a consequence, once a spin state and spin labels have been specified, 
the equation above can be rewritten as:
\begin{equation}
\frac{\int dR \psi_T^*(R) \psi_T(R) \frac{O \psi_T(R)}{\psi_T(R)}}{\int dR \psi_T^*(R) \psi_T(R)},
\end{equation}
where $R$ here represents the space coordinates for the system.
\newline
On the other hand, for Hamiltonians containing a spin-orbit interactions, the same 
simplification is not correct.
The trial wave function will be, in general, made up with Slater determinants containing single 
particle spinors and all the spin variables should be in this case taken into account explicitly. 
Spinors in general have nontrivial forms, and the spin state of the single particles could change
as a function of the space coordinates.
This means that in general one needs to perform a summation over a complete spin basis:
\begin{equation}
\int dR \sum_S\frac{<\psi_T|R,S><R,S|O|\psi_T>}{<\psi_T|\psi_T>}.
\label{evspin}
\end{equation}
Clearly, one way or another, one has to sample the relevant degrees of freedom in 
order to evaluate the impact of the spin on the expectation values. One possible way is
to sum over the spin variables, {\it i.e.} sample the $2^N$ discrete space.
\newline
What is proposed here is an alternative approach in which integration 
takes the place of the discrete sum, and by exploiting the VMC algorithm this can be performed 
with good computational efficiency.
\newline
The trial wave function which we employ in calculations is 
a single Slater determinant of single-particle spinors.
Let us first consider the simple  two-electron case in order to illustrate the method.
The spinors we will use are
\begin{equation}
\vec{\psi}_1 = \left(
  \begin{array}{c}
      \psi^1_1 \\
      \psi^1_2 \\
  \end{array} \right)
\qquad 
\vec{\psi}_2 = \left(
  \begin{array}{c}
      \psi^2_1 \\
      \psi^2_2 \\
  \end{array} \right)
\end{equation}
The spacial coordinates of the two electrons will be $\vec{r_1}$ and $\vec{r_2}$ while the spin
coordinates can be parameterized as follows:
\begin{equation}
\vec{s}_1 = \left(
  \begin{array}{c}
      \sin(\alpha_1) e^{i\delta_1} \\
      \cos(\alpha_1) \\
  \end{array} \right)
\qquad 
\vec{s}_2 = \left(
  \begin{array}{c}
      \sin(\alpha_2) e^{i\delta_2} \\
      \cos(\alpha_2) \\
  \end{array} \right)
\label{spinalf}
\end{equation}

The Slater determinant can be written as
\begin{equation}
\Phi(\vec{r}_1,\vec{s}_1,\vec{r}_2,\vec{s}_2) = {\rm det} \left(
  \begin{array}{cc}
      <\vec{s}_1,\vec{r}_1|\vec{\psi}_1>  & <\vec{s}_1,\vec{r}_1|\vec{\psi}_2>   \\
      <\vec{s}_2,\vec{r}_2|\vec{\psi}_1>  & <\vec{s}_2,\vec{r}_2|\vec{\psi}_2>   \\
  \end{array} \right).
\end{equation}
A complete spin basis for this system is $\uparrow \uparrow$,$\uparrow \downarrow$,$\downarrow \uparrow$,$\downarrow \downarrow$, so that, given a second wave function $\Phi'$ of the same form as $\Phi$, the following equality holds:
\begin{eqnarray}
<\Phi|\Phi'>=<\Phi|\uparrow \uparrow><\uparrow \uparrow|\Phi'>+
<\Phi|\uparrow \downarrow><\uparrow \downarrow|\Phi'> \nonumber \\
+<\Phi|\downarrow \uparrow><\downarrow \uparrow|\Phi'>+
<\Phi|\downarrow \downarrow><\downarrow \downarrow|\Phi'>. \qquad
\label{spinsumm}
\end{eqnarray}
The terms  involved in this summation are:
\begin{eqnarray}
<\uparrow \uparrow|\Phi>=\psi^1_1(\vec{r_1})\psi^2_1(\vec{r_2})-\psi^2_1(\vec{r_1})\psi^1_1(\vec{r_2}) \nonumber \\
<\uparrow \downarrow|\Phi>=\psi^1_1(\vec{r_1})\psi^2_2(\vec{r_2})-\psi^2_1(\vec{r_1})\psi^1_2(\vec{r_2}) \nonumber \\
<\downarrow \uparrow|\Phi>=\psi^1_2(\vec{r_1})\psi^2_1(\vec{r_2})-\psi^2_2(\vec{r_1})\psi^1_1(\vec{r_2}) \nonumber \\
<\downarrow \downarrow|\Phi>=\psi^1_2(\vec{r_1})\psi^2_2(\vec{r_2})-\psi^2_2(\vec{r_1})\psi^1_2(\vec{r_2}) \,.
\label{sterms}
\end{eqnarray}
Writing $\Phi$ explicitly as
\begin{eqnarray}
\Phi(\vec{r}_1,\vec{s}_1,\vec{r}_2,\vec{s}_2) = \qquad \qquad \qquad \nonumber \\
=\left(\psi^1_1(\vec{r}_1)e^{i\delta_1}\sin(\alpha_1)+\psi^1_2(\vec{r}_1)\cos(\alpha_1)\right) \cdot \nonumber \\
\left(\psi^2_1(\vec{r}_2)e^{i\delta_2}\sin(\alpha_2)+\psi^2_2(\vec{r}_2)\cos(\alpha_2)\right) \nonumber \\
-\left(\psi^2_1(\vec{r}_1)e^{i\delta_1}\sin(\alpha_1)+\psi^2_2(\vec{r}_1)\cos(\alpha_1)\right) \cdot \nonumber \\
\left(\psi^1_1(\vec{r}_2)e^{i\delta_2}\sin(\alpha_2)+\psi^1_2(\vec{r}_2)\cos(\alpha_2)\right)
\end{eqnarray}
it is possible to demonstrate how the summation \eqref{spinsumm} can be exactly
rewritten as an integral over the spin parameters $\alpha_{1,2}$ and $\delta_{1,2}$ in the following
way:
\begin{eqnarray}
<\Phi|\Phi'>=Const \cdot \int \Big[ \int_0^{2\pi} \Phi(\vec{r}_1,\vec{s}_1,\vec{r}_2,\vec{s}_2)^* \Phi'(\vec{r}_1,\vec{s}_1,\vec{r}_2,\vec{s}_2) \nonumber \\
d\alpha_1 d \alpha_2 d \delta_1 d \delta_2 \Big] d \vec{r}_1 d\vec{r}_2 \qquad 
\label{sumint}
\end{eqnarray}
In order to prove the above statement we take into consideration the terms corresponding to
$<\Phi|\uparrow \uparrow>$$<\uparrow \uparrow|\Phi'>$, since all other terms can be obtained in the same way.
First of all we notice that
\begin{eqnarray}
\int_0^{2\pi} d \alpha_i \int_0^{2\pi} d \alpha_j \sin(\alpha_i) \sin(\alpha_j) = \delta_{i,j} 2\pi^2 \nonumber \\
\int_0^{2\pi} d \alpha_i \int_0^{2\pi} d \alpha_j \cos(\alpha_i) \cos(\alpha_j) = \delta_{i,j} 2\pi^2 \nonumber \\
\int_0^{2\pi} d \alpha_i \int_0^{2\pi} d \alpha_j \sin(\alpha_i) \cos(\alpha_j) = 0 .
\label{sincos}
\end{eqnarray}
This means that only those terms with pairs of sines and cosines of the same angles will give
a non zero contribution.
Using these relations we can select only the interesting terms:
\begin{eqnarray}
\int_0^{2\pi} d \alpha_1 d \alpha_2 \big[ \psi^1_1(\vec{r}_1)^*e^{-i\delta_1}\sin(\alpha_1)  \psi^2_1(\vec{r}_2)^* e^{-i\delta_2} \sin(\alpha_2) \nonumber \\
-\psi^2_1(\vec{r}_1)^*e^{-i\delta_1}\sin(\alpha_1)  \psi^1_1(\vec{r}_2)^*e^{-i\delta_2} \sin(\alpha_2) \big] \nonumber \\
\cdot \big[ \psi'^1_1(\vec{r}_1)e^{i\delta_1}\sin(\alpha_1)  \psi'^2_1(\vec{r}_2)e^{i\delta_2} \sin(\alpha_2) \nonumber \\
-\psi'^2_1(\vec{r}_1)e^{i\delta_1}\sin(\alpha_1)  \psi'^1_1(\vec{r}_2)e^{i\delta_2}\sin(\alpha_2) \big] \qquad
\end{eqnarray}
which give
\begin{eqnarray}
Const \cdot \big[ \psi^1_1(\vec{r}_1)^* \psi^2_1(\vec{r}_2)^* \psi'^1_1(\vec{r}_1) \psi'^2_1(\vec{r}_2)  \nonumber \\
+\psi^2_1(\vec{r}_1)^* \psi^1_1(\vec{r}_2)^* \psi'^2_1(\vec{r}_1) \psi'^1_1(\vec{r}_2) \qquad \nonumber \\
-\psi^1_1(\vec{r}_1)^* \psi^2_1(\vec{r}_2)^* \psi'^2_1(\vec{r}_1) \psi'^1_1(\vec{r}_2) \qquad \nonumber \\
-\psi^2_1(\vec{r}_1)^* \psi^1_1(\vec{r}_2)^* \psi'^1_1(\vec{r}_1) \psi'^2_1(\vec{r}_2) \big], \qquad
\end{eqnarray}
and this is precisely  what one would obtain from the products of the first row of Eq. \eqref{sterms}.
As mentioned above, all the other terms of \eqref{spinsumm} can be obtained in the same way, proving
the equality Eq. \eqref{sumint}. Furthermore, one can easily see how the integration over the variables
$\delta_i$ is not necessary since these variables appear only as a phase, which exactly cancels out
in the non-vanishing terms.
Besides this, though the proof of \eqref{sumint} was explicitly given for the particular $N=2$ case, it can 
be shown that the same relation holds for any $N$. This property holds due to
the relations \eqref{sincos}, which actually ensure a correct selection of all the necessary 
terms. The basic idea is that up (down) states of any particle will only be matched to the
corresponding up (down) states of the same particle, and this precisely corresponds to projection over a fixed
spin state. Since all combinations are taken into account by the presence of both up and down
states for all particles, we have accomplished 
the summation over the entire basis set.
\newline
All interesting observables $O$ which can be estimated with VMC, when acting on a Slater determinant, require at most the use of a linear combination of some new Slater 
determinants. This can be expressed as
\begin{equation}
<R,S|O|\Phi>\sum_i <R,S|\Phi'_i>
\end{equation}
and for linearity the expressions we found can easily be applied.
At this point, equation \eqref{evspin} can be rewritten as
\begin{eqnarray}
\frac{\int dR d{\alpha} \psi_T^*(R,\alpha) \psi_T(R,\alpha) \frac{O \psi_T(R,\alpha)}{\psi_T(R,\alpha)}}{\int dR d\alpha \psi_T^*(R,\alpha) \psi_T(R,\alpha)}.
\end{eqnarray}
where the spin coordinates of the system $S$ have been rewritten in terms of the parameters set
${\alpha}$ using \eqref{spinalf}.
\newline
As mentioned earlier, the proposed algorithm does not represent the only possibility for
spin summation. A possible alternative could be that of a VMC-sampled sum over
the possible spin states and this would also represent an efficient algorithm.

\section{Applications}
A broad class of problems exists in which the method discussed above can be applied.
Nevertheless, it must also be stressed that it is not always easy to find a good trial wave
function when an interaction of the spin-orbit kind is included in the Hamiltonian. This
is mainly due to the non-locality of the spin-orbit interactions.
\newline
In order to test the applicability of the method to atomic systems in presence of spin-orbit 
interactions, we chose to test this method on isolated C, Tl, Pb, Bi and Po atoms. 
By investigating both a light atom like Carbon and heavier elements, diverse spectra showing
very different energy splittings are considered,
corresponding to the two distinct limits of LS and jj coupling.

\subsection{Carbon atom}
The carbon atom has six electrons and can be considered as a light element.
Although the spin-orbit effects are known to be very small and of little relevance for
most properties,
they still induce observable splittings in the energy spectrum.
Furthermore, the computation of these very small energy differences represents a
good starting test for the efficiency of the method.
The all-electron
Hamiltonian which was employed in the calculations concerning the carbon atom is
given by
\begin{equation}
		H=\sum_{i=1}\left( \frac{\mathbf{P}_i^2}{2m} +V_{SO}^i -
 \frac{Z}{\mathbf{r}_i}\right)+\sum_{i<j}\frac{1}{\mathbf{r}_{ij}}
\label{ham}
\end{equation}
where $V^i_{SO}$ accounts for the spin-orbit interaction for the $i-th$ electron. 
It should be noted that the spin-orbit potential is the only relativistic effect 
contained in this Hamiltonian. The rest of the relativistic corrections 
were neglected since they do not contribute to the level splittings 
if we assume the wave functions given below.
The trial wave functions employed in the calculations were linear combinations of Slater
determinants of single particle Hartree-Fock orbitals \cite{clementiroetti}.
Since the interest, regarding the C atom, was focused on providing a good test for the method 
using a well-known type of wave function, no optimization was performed, 
and the  combinations of Slater determinants were fixed by the imposed 
spin-spatial symmetries in the LS coupling.
\newline
The spin-orbit interaction in atoms comes from an approximated form of the Dirac equation and
 can be written (for the $i-th$ particle) as
\begin{equation}
V_{SO}^i \psi = -\frac{\hbar^2}{4m^2c^2}i\vec{\sigma}_i\cdot\left[ \left( \vec{\bigtriangledown}_iV \right) 
\times \vec{\bigtriangledown}_i \psi \right]
\label{VSO}
\end{equation}
where $V$ in this case is the total potential felt by the considered particle. This potential 
contains both the attractive contribution from the nucleus and the sum of the repulsive 
interactions with all other electrons. What is commonly done is approximating $V$ with
an effective potential $V_{eff}$ having spherical symmetry. Within this approximation, equation 
\eqref{VSO} can be rewritten as
\begin{equation}
\frac{1}{2m^2c^2}\frac{1}{r}\frac{dV_{eff}^i}{dr}\vec{L}_i\cdot\vec{S}_i \psi.
\label{effso}
\end{equation}
In case the Coulomb repulsion among electrons is neglected, one is left with a sum of single
particle Hamiltonians and if the factor in Eq. \eqref{effso} multiplying $\vec{L}_i\cdot\vec{S}_i$ 
is substituted
by a constant, the problem can be exactly solved in analytic form.
The term $\vec{L}_i\cdot\vec{S}_i$ in fact commutes with the rest of the Hamiltonian, which is spherically symmetric.
However, as soon as the electron-electron Coulomb interaction is taken into account the exact 
solution is unknown. 
As already mentioned, since the SO interaction in C is known to yield very small energy
splittings, it is reasonable to treat it within the LS coupling procedure.
This is done by writing the Hamiltonian \eqref{ham}
as $H=H_1+H_2$ where $H_2=\sum_i V_{SO}^i$ and combining eigenstates of $H_1$ (which are
 eigenstates of both $\mathbf{L}$ and $\mathbf{S}$) in order to obtain eigenstates of $\mathbf{J}$.
Calling the eigenstates of $H_1$ as $|\gamma L S M_L M_S>$ ($\gamma$ stands for all residual
 quantum numbers), it may be shown by using the Wigner-Eckart theorem that
\begin{eqnarray}
<\gamma L S M_{L} M_{S}| H_{2} |\gamma L S M_{L}' M_{S}'>= \nonumber \qquad\\
{\cal{A}} <\gamma L S M_L M_S|\mathbf{L}\cdot\mathbf{S}|\gamma L S M_L' M_S'>
\end{eqnarray}
where $\cal{A}$ is a constant depending on $(\gamma L S)$.
The eigenstates of $\mathbf{J}$ can be obtained as linear combinations of the $|\gamma L S M_L M_S>$
states and can be denoted as $|\gamma L S J M_J>$.
At this point it is easily shown that
\begin{eqnarray}
<\gamma L S J  M_{J}| H_{2} |\gamma L S J M_{J}>= \nonumber \qquad\\
\frac{1}{2}{\cal{A}} \left[ J(J+1)-L(L+1)-S(S+1)\right]
\end{eqnarray}
Notice how the expectation value of a sum of single particle terms $H_2$ is related to that of 
$\mathbf{L}\cdot\mathbf{S}$ ($\mathbf{S}$ and $\mathbf{L}$ are the total spin and angular
 momentum) by the constant $\cal{A}$. This constant contains the effect of the effective potential $V_{eff}$
and in general it cannot be determined analytically.
However, when the SO interaction \eqref{VSO} is approximated by $\cal{C}$$\mathbf{L}_i\cdot\mathbf{S}_i$ 
with $\cal{C}$ constant, $\cal{A}$ can be calculated exactly.
A first test for our method consisted in verifying the relations above and comparing the numerical
result for $\cal{A}$ with the analytically computed value.
In table \ref{tabellaen} numerical and analytical results are compared for three different states,
 showing excellent agreement with the predictions of the Wigner-Eckart theorem.

\begin{table}[ht]
\begin{tabular}{ccccccc}
                                                                                                    
\multicolumn{7}{c}{} \\
\hline
\hline
L & S & J  & numerical  & analytical& numerical & analytical \\
  &   &   & $<\sum_i\mathbf{L}_i\cdot\mathbf{S}_i>$  &  $<\sum_i\mathbf{L}_i\cdot\mathbf{S}_i>$  & ${\cal{A}}$ & ${\cal{A}}$ \\
\hline
1 & 1 & 0 & -1.001(2)  & -1.0  & 0.5005(1)  & 0.5     \\
1 & 1 & 1 & -0.5004(7) & -0.5  & 0.5004(7)  & 0.5     \\
1 & 1 & 2 &  0.4997(5) &  0.5  & 0.4997(5)  & 0.5     \\
\hline
\hline
\end{tabular}
\caption{Analytical and numerical results for the mean value of the SO interaction ($\cal{C}$=1) and effective coupling constant $\cal{A}$ over the three lowest energy states given by fixed $L=1$,$S=1$ and $J=0,1,2$. Results are reported in atomic units}
\label{tabellaen}
\end{table}

A second set of calculations was then performed, including this time a SO interaction of the form 
\eqref{effso} in order to obtain an estimate of the energy splitting induced by the effective
spherical potential $V_{eff}$. 
The only contributions to the SO splitting in the C atom come from the two occupied $2p$ 
orbitals, therefore it was only necessary to take the effective potential $V_{eff}$ for $2p$
electrons into account.
Following Slater \cite{slater}, in a single particle picture one could write the effective potential for the particle $i$ as
\begin{eqnarray}
V_{eff}^i(r)=\frac{Z_{eff}^i(r)}{r} \nonumber \\
Z_{eff}^i(r)=Z-\int_0^{r}\sum_{j\neq i} |\phi_j(\mathbf{r})|^2 d\mathbf{r}
\end{eqnarray}
VMC calculations were done for the energy splittings of the lowest energy states ($L=1,S=1$).
The results (see table \ref{tabv}) are in qualitatively good agreement with experimental data \cite{expc}.
The relative energy differences between ($J=0$,$J=1$) and ($J=1$,$J=2$) states follow the theoretical
ratio ($1/2$), though they both are too large. The fact that splittings from experimental
data do not show the theoretical $1/2$ ratio makes the result for the larger splitting worse.
In order to check whether our VMC method
 would give a correct description also in case of a realistic
SO interaction, further testing was done: in case $V_{eff}$ is substituted by $Z/r^2$ and the
radial parts of the $2p$ single particle wave functions are given by a single orbital of the Slater type
\begin{eqnarray}
S_{jl}(r)=N_{jl}r^{n_{jl}-1}\exp({-z_{jl}r}) \nonumber \\
N_{jl}=(2z_{jl})^{n_{jl}+1/2}/[(2n_{jl})!]^{1/2}
\end{eqnarray}
($n_{ij}=1,2,3,..$ and $z_{ij}$ are fitting parameters) one can obtain an exact analytical form
for the SO splitting. Also in this case the numerical results were in excellent agreement with
the theoretical predictions.

\begin{table}[ht]
\begin{tabular}{ccccccc}
\multicolumn{7}{c}{} \\
\hline
\hline
L & S & J & & realistic $V_{SO}$	& & effective $V_{SO}$		 \\
\hline
1 & 1 & 0 & & -2.9(2)  $\cdot 10^{-5}$	& & -3.3(2)   $\cdot 10^{-5}$    \\
1 & 1 & 1 & & -1.4(1)  $\cdot 10^{-5}$	& & -1.6(1)   $\cdot 10^{-5}$    \\
1 & 1 & 2 & &  1.4(1)  $\cdot 10^{-5}$	& &  1.7(2)   $\cdot 10^{-5}$    \\
\hline
\hline
\end{tabular}
\caption{Numerical results for the mean value of the realistic \eqref{VSO} and effective SO interaction \eqref{effso} per electron for the three lowest energy states. Results are reported in atomic units}
\label{tabv}
\end{table}

In order to check the effects of the approximations contained in the
effective spherical potential, calculations were done substituting \eqref{effso} in the Hamiltonian
with the realistic SO interaction \eqref{VSO}. Once more, SO potential expectation values were computed for the three lowest energy levels. In this case the energy splittings are reduced
with respect to the previous case becoming closer to experimental data (see Table III).
The relative spacings retain the theoretical ratio $1/2$ predicted for spherical potentials as
before, but the 
effective SO coupling constant is close to the experimental one, at least for the smaller 
splitting ($J=0$, $J=1$). It remains however about $30$ percent too large in the ($J=1$, $J=2$) 
case.
\newline
In the first case the effective screening of the inner electrons appeared to be not effective enough
and one could be led to the conclusion that Hartree-Fock (HF) orbitals do not provide an accurate charge 
distribution. However, this last calculation showed how, at least for the lowest splitting,
the screening effects affecting the SO coupling are reasonably well reproduced in the C
atom already at the HF level.

\begin{table}[ht]
\begin{tabular}{cccc}
                                                                                                    
\multicolumn{4}{c}{} \\
\hline
\hline
 J & VMC (effective $V_{SO}$)    & VMC (realistic $V_{SO}$)    &  experimental	     \\ 
\hline
 0 & $-10(1.5)\cdot10^{-5}$  & $ -9(1.5)\cdot10^{-5}$  &   $ -7.5\cdot10^{-5}$          \\
 1 & 0.0  	      &  0.0      	  &   0.0       		\\
 2 & $20(1)\cdot10^{-5}$  &  $17(1)\cdot10^{-5}$  &   $ 12.3\cdot10^{-5} $          \\
\hline
\end{tabular}
\caption{All electrons SO splittings (defined as $E(J)-E(1)$) for C. In the second and third columns VMC predictions relative to effective
SO and realistic SO respectively. Experimental values in the fourth columns.  Results are reported in atomic units}
\label{pbsplit}
\end{table}

\subsection{Tl, Pb, Bi and Po atoms}
In order to give a more complete picture of the applicability of the algorithm, the VMC method
was also applied to the Tl (Z=81), Pb (Z=82), Bi (Z=83) and Po (Z=84) atoms, which exhibit sizable SO interaction effects.
While for C the LS coupling was employed, due to the increased strength of the SO interaction,  
a better approximation for the description of the energy splittings in heavier atoms
is given by the $jj$ coupling scheme.
For this reason, the wave function is obtained by combining single-particle $J_i$ (for the $i-th$ 
particle) states into the eigenstates of the total angular momentum $J$.
\newline
Due to the large number of electrons in the Tl to Po atoms, and the very deep ionic potential, 
it is customary to limit the explicit degrees of freedom to valence electrons by introducing an effective 
description in terms of pseudopotentials (or effective core potentials).
Although there is no  a-priori limitation in the number of electron that can be managed in a VMC calculation, very large energy fluctuations  coming from the core states are very expensive to average out and are of marginal
interest in determining the quantities we are interested in, given that SO splitting is only related to the presence of
occupied valence $p$-states.
The  pseudopotential employed was taken from Kuechle et.al. \cite{pseudo},
leading to a Hamiltonian (in atomic units) of the following form
\begin{equation}
-(1/2)\sum_i {\mathbf \bigtriangledown}^2_i+V_{pp}+\sum_{i<j}\frac{1}{r_{ij}},
\end{equation}
where the indices $i,j$ run over the valence electrons and $V_{pp}$ is the sum of a spin-orbit
averaged ab initio pseudopotential $V_{av}$, an SO operator $V_{so}$ and a term representing
effective charge of the pseudo-nucleus.
\newline
The use of this pseudopotential has the additional benefit of making possible the comparison to the HF
results reported in the same reference.
In our calculations, the radial parts of  $s$ and $p$ orbitals were represented by an
expansion in four uncontracted gaussians.
The single-particle orbitals optimization was performed with a variational approach using an
adapted implementation of the FIRE algorithm \cite{fire}. This simple algorithm is based on a
molecular dynamics approach to the search in the variational parameters space. The parameter
evolution is driven by the forces induced by
the total energy partial derivatives while appropriate cooling is introduced as
parameters motion is stopped whenever the velocities directions are opposite to the
forces.
The forces, in turn, are computed during the VMC runs through a sampling of the wave function 
derivatives with respect to the parameters. Given the form of the trial wave 
function $\psi_T$, the corresponding differentiation of the energy expectation
 according to a parameter $\alpha$  
is given by 
\begin{equation}
\partial_{\alpha}E_{\psi_T}=\partial_{\alpha}\frac{<\psi_T |H|\psi_T >}{<\psi_T|\psi_T>}
\end{equation}
and can be evaluated
 following the method also employed by Sorella\cite{sorella,sorella2}, as 
\begin{equation}
-2\frac{<\psi_T|H|\psi_T><\psi_T|\partial_{\alpha}|\psi_T>}{(<\psi_T|\psi_T>)^2}
+2\frac{<\psi_T|H\partial_{\alpha}|\psi_T>}{<\psi_T|\psi_T>}.
\end{equation}
Given the physical relevance of SO splittings in the heavy atoms here considered, the influence of
correlation effects was investigated by introducing in the wave function a Jastrow factor
\cite{mitasrev} of the Pade' form:
\begin{eqnarray}
J(\mathbf{R})=exp\left(\sum_{i,j}^N V(r_{ij}) \right) \nonumber \\
V(ri_{ij})=-\frac{r_{ij}}{4(1+ar_{ij})}
\end{eqnarray}
being $a$ a variational parameter.
This factor introduces two body correlations, which in VMC calculations are usually necessary to
obtain a realistic description of the correlation energy.
In principle, the Jastrow factor should depend on the relative spin state of the electron pair. In presence
of SO interactions inducing a spin rotation,  this requirement would lead to a correlated wave function of the 
form $\prod_{i,j}f_{\sigma\sigma}(r_{ij})\vec{\sigma}_{i}\cdot\vec{\sigma}_{j}$, which would in turn imply the necessity of computing a 
sum over all the possible two electron spin states. This sum grows as $2^{N}$, and becomes very quickly 
unmanageable. For this reason we prefer, for the moment, to completely neglect  the spin dependence
in the two-body wave function, focusing on the gross effect of introducing the short range correlations
induced by the Coulomb potential.
We stress, however, that the introduction of correlation effects already implies a huge improvement with
respect to the HF method, keeping the computational cost at a reasonable level.
\newline
In computing the SO splittings, the lowest  
states corresponding to the total angular momentum $J$ eigenvalues, obtained from
the multiple occupation of single particle $p$ orbitals were considered for each atom.
Comparing for instance Pb and C, despite the equal number of valence electrons, due to the diverse
couplings ($jj$ and $LS$, respectively), these show different splitting patterns.
In particular, for Pb the $(J=0,J=1)$ splitting is larger than the difference between $(J=1,J=2)$.
\newline
VMC results  are reported in table \ref{pbsplit} together with the HF results of
Kuechle \cite{pseudo} and the corresponding experimental values \cite{sperpb}.
Calculations were also performed in absence of correlation, by removing the jastrow factor.
Since in this case the trial function is given by the Slater determinant of spinors 
we can compare our results with the corresponding HF calculation. A good compatibility is found.
Interestingly, correlation effects on the SO splittings appear to be smaller than the
statistical error, suggesting that SO effects are mainly determined by the single particle
properties of the wave function.
From the comparison among the different data sets we confirm that our VMC calculation
 shows the correct ordering of the 
SO splittings among the states considered. Indeed, our energy differences agree
with the other two data sets essentially within
the statistical error bars.

\begin{table}[ht]
\begin{tabular}{cccccc}
                                                                                                    
\multicolumn{6}{c}{} \\
\hline
\hline
  &  J & VMC uncorrelated & VMC+jastrow   & HF	    & experimental     \\
\hline
Tl&    &            &           &           &        \\
  &1/2 & 0.0	    &  0.0      & 0.0       & 0.0       \\
  &3/2 & 0.030(5)   &  0.031(5) & 0.033     & 0.035     \\
\hline
Pb&    &            &           &           &        \\
  &  0 & 0.0        &  0.0      & 0.0       & 0.0       \\
  &  1 & 0.032(4)   &  0.033(4) & 0.029     & 0.035     \\
  &  2 & 0.044(5)   &  0.045(5) & 0.046     & 0.048     \\
\hline
Bi&    &            &           &           &        \\
  &3/2 & 0.0	    &  0.0      & 0.0	    & 0.0       \\
  &3/2 & 0.054(4)   &  0.048(6) & 0.057     & 0.052     \\
  &5/2 & 0.074(5)   &  0.070(4) & 0.078	    & 0.070     \\
\hline
Po&    &            &           &           &        \\
  &  2 & 0.0	    &  0.0      & 0.0	    & 0.0       \\
  &  0 & 0.049(8)   &  0.043(8) & 0.042	    & 0.034     \\
  &  1 & 0.080(6)   &  0.077(5) & 0.071     & 0.076     \\
\hline
\hline
\end{tabular}
\caption{SO splittings for Tl, Pb, Bi and Po. The second and third columns contain uncorrelated
and correlated VMC results respectively. The fourth and fifth columns report HF results by Kuechle
and the relative experimental values \cite{sperpb}. States associated with zero splitting values are taken as a reference.
Results are reported in atomic units}
\label{pbsplit}
\end{table}

\section{Conclusions}
We introduced an extension of the VMC method capable to realistically describe spin-orbit splittings
in heavy atoms. Calculations were tested first in the light C atom, and then extended to a set
of heavier open $p$-shell atoms (Ti to Po).
We demonstrated that the algorithm  provides correct evaluation of the underlying integrals and, remarkably, leads to sufficient 
accuracy for resolving the small SO energy differences. 
\newline
For the case of the carbon atom
 the small SO splitting is already reasonably well captured within a non-relativistic
Hartree-Fock wave function, although an even more accurate description would require the use of a better variational
wave function. Interestingly, the investigation of the effects of the spherical effective SO term showed
an appreciable improvement when the more realistic version of the SO coupling operator was 
employed.
\newline
Calculations of SO energy
splittings carried on in heavier atoms required a wave functions based on $jj$ rather than
$LS$ coupling. In these systems the possibility of a direct optimization of wave functions
was investigated.  The effects of electron-electron correlations have been studied by introducing
in the wave function an explicit, though simplified, two-body Jastrow factor. At this level dynamical
quantum correlations seem not to have a huge effect on the SO splittings estimation.
\newline
We believe that the obtained results are encouraging and the method can be applied 
to more complex systems while retaining the efficiency and robustness of the VMC 
implementation.

\section{Acknowledgments}
Calculations were performed on the Wiglaf cluster at the Physics Department, University of Trento.
 Research of one of us (L.M.) 
is supported by the DOE LANL subcontract 81279-001-10, by  NSF grants 
DMR-0804549 and OCI-0904794 and by ARO.

\end{document}